\documentclass[12pt]{article} 
\usepackage{graphicx}
\begin{document}
\baselineskip 10mm

\centerline{\bf Single-qubit operations in the double-donor
structure} \centerline{\bf driven by strongly detuned optical
pulses.} \vskip 2mm

\centerline{Alexander V. Tsukanov}

\vskip 2mm

\centerline{\it Institute of Physics and Technology, Russian
Academy of Sciences}
\centerline{\it Nakhimovsky pr. 34, Moscow
117218, Russia}
\centerline{\it E-mail: tsukanov@ftian.oivta.ru}

\vskip 4mm

\begin{abstract}

We study theoretically the quantum dynamics of an electron in the
singly-ionized double-donor structure in the semiconductor host
under the influence of two strongly detuned laser pulses. This
structure can be used as a charge qubit where the logical states
are defined by the lowest two energy states of the remaining
valence electron localized around one or another donor. The
quantum operations are performed via Raman-like transitions
between the localized (qubit) states and the manyfold of states
delocalized over the structure. The possibility of realization of
arbitrary single-qubit rotations, including the phase gate, the
NOT gate, and the Hadamard gate, is demonstrated. The advantages
of the off-resonant driving scheme for charge qubit manipulations
are discussed in comparison with the resonant scheme proposed
earlier.

\end{abstract}

\vskip 4mm

PACS number(s): 03.67.Lx, 78.67.-n

\vskip 6mm

\newpage

\centerline{\bf I. INTRODUCTION}

  In view of recent progress in the development of the controlled-donor implantation techniques \cite{1},
Kane's paradigm of  the solid-state quantum computations \cite{2}
has gained new insights. The alternative schemes using the orbital
\cite{3,4} or spin \cite{5,6} degrees of freedom of the
donor-bounded electrons to encode the quantum information instead
of the nuclear donor spins, have been proposed. Besides, several
refinements of the original proposal concerning the initialization
\cite{7} and read-out \cite{8,9,10,11} as well as the information
transfer through the quantum networks \cite{12,13,14} have been
made.

In particular, the pair of donors sharing an electron has been
considered as very promising candidate for the solid-state qubit
embodiment \cite{3}. The qubit is presented by the electron
orbital states positioned at the different donors. There are two
main driving mechanisms for the coherent electron evolution
defining a quantum operation on such a qubit. First uses the
electric fields through an application of the adiabatically
switched voltages to the surface gates placed above the donor
structure to modify the confinement potential, thus varying the
electron tunnelling rates between neighboring donors \cite{3,15}.
The desired state of the qubit is realized by an appropriate
choice of the voltage parameters. The second scheme relies upon
the optical dipole transitions between the size-quantized
one-electron levels induced by the resonant pulses \cite{16,17}.
As it was shown, an arbitrary single-qubit operation can be
achieved with two simultaneously switched pulses connecting the
qubit states via the excitation of the state delocalized over the
structure. The latter scheme is likely to be more preferable than
the former due to its higher selectivity and lower field
intensity. The successful implementation of the quantum
operations, however, requires one to provide the high precision in
the durations, the frequencies, the polarizations and the
strengths of the pulses together with reliable control over the
delay time and the phase difference between the pulses. Besides,
the use of the intermediate state lying in the neighborhood of the
continuum introduces the decoherence caused by the ionization and
the spontaneous emission
 from this state.

Here we propose the way to overcome the difficulties inherent to
the electron charge manipulations by optical means. We show that
the off-resonant laser pulses can be used to generate an arbitrary
rotation of the qubit-state vector as well as to drive an electron
between the remote donors. It is essential that this method is
based on the Raman-like transitions between the localized electron
states of an effective molecular ion where the excited
(delocalized over the double-donor system) states irrespective of
their number are used as the transport channels. As we shall see,
the coherent electron dynamics is described by the simple
analytical model. The pulse and structure parameters needed for
that type of quantum evolution may be evaluated from the results
obtained in this study.

\centerline{\bf II. MODEL}

We begin with the description of the one-electron double-donor
(DD) structure (Fig. 1). Let the donors $A$ and $B$ be placed on
the axis $x$ from each other at the distance $R$ large enough to
consider their ground orbital states, $\left| {A0} \right\rangle $
and $\left| {B0} \right\rangle $, to be isolated. Due to this fact
those states may be used as the qubit states $\left| {0}
\right\rangle $ and $\left| {1} \right\rangle $, respectively (if
an electron is localized on the donor $A (B)$, the qubit is in the
state $\left| {0} \right\rangle $ ($\left| {1} \right\rangle $)).
The energy difference $\Delta  = \varepsilon _{B0}  - \varepsilon
_{A0} \equiv \varepsilon _1  - \varepsilon _0 $ may be introduced
due to the structure asymmetry caused by the fabrication process
and/or by the bias voltages $V$ applied to the donors. The
coupling between the excited states of the donors through the
electron tunnelling gives rise to the forming of hybridized states
delocalized over the DD structure. However, if $R\gg a_B$ ($a_B$
is the effective Bohr radius of host material) the low-lying
excited states are hybridized weakly and do not participate the
two-donor dynamics. We will consider therefore only the excited
states of individual donors whose orbitales considerably overlap,
e.g., the $|nP_x\rangle$ states. The resulting effective
single-electron spectrum of the DD structure is presented by the
sequence of the states $\left\{ {\left| {k} \right\rangle }
\right\}_{exc}$ which for $\Delta=0$ are the doublets composed of
the symmetric and antisymmetric superpositions of isolated donor
states. If $\Delta\neq0$, the spectrum is expected to be much more
complex. Taking into account the hydrogen-like spectrum of the
isolated donors we expect the excited states close to the edge of
the potential barrier separating the donors to have the
quasi-continuous energy distribution. The energy gap $\omega
_{exc}  = \varepsilon _{exc}^{\min }  - \varepsilon _1 $ between
the state $\left| {1} \right\rangle $ and the lowest state
${\left| {k } \right\rangle }_{exc}^{\min }$ from that manyfold is
assumed to be greater than all other energy scales relevant for
consideration:
\begin{equation}
\omega _{exc}  > \Delta ,\,\,\,\omega _{exc}  > \max \Delta _{mn}
,\,\,\Delta _{mn}  = \left| {\varepsilon _m  - \varepsilon _n }
\right|,\,\,m,n \in \left\{ {k} \right\}_{exc}.
\end{equation}

 In what follows we shall study the one-electron
quantum dynamics involving the localized (qubit) states, $\left|
{0} \right\rangle $ and $\left| {1} \right\rangle $, and the
states $\left\{ {\left| {k} \right\rangle } \right\}_{exc}$
delocalized over the structure. Our aim is to choose the field and
structure parameters so that to drive an initial qubit state
$\left| {\Psi \left( t_0 \right)} \right\rangle  = \alpha _0
\left| 0 \right\rangle  + \beta _0 \left| 1 \right\rangle =
\left(\alpha_0, \beta_0 \right)^T$ into the final state $\left|
{\Psi \left( t \right)} \right\rangle  = \alpha \left| 0
\right\rangle + \beta \left| 1 \right\rangle = \left(\alpha, \beta
\right)^T$ with the desired coefficients $\alpha$ and $\beta $.

In the absence of an external field the DD structure is
characterized by the stationary Hamiltonian $H_0$ with the
eigenstates $\left\{ {\left| k \right\rangle } \right\}$ and the
eigenenergies $\left\{ {\varepsilon _k } \right\}$:
\begin{equation}
H_0 \left| k \right\rangle  = \varepsilon _k \left| k
\right\rangle.
\end{equation}
The eigenstates $\left\{ {\left| k \right\rangle } \right\}$ form
the complete orthonormal set so that
\begin{equation}
\sum\limits_k {\left| k \right\rangle \left\langle k \right|}  =
1.
\end{equation}
In the presence of the electromagnetic field the system
Hamiltonian reads
\begin{equation}
H = H_0  - e{\bf{E}}\left( t \right){\bf{r}},
\end{equation}
where $e$ is the electron charge, ${\bf{E}}\left( t \right)$ is
the field strength, $\bf{r}$ is the radius-vector of an electron.
With the help of Eqs. (2) and (3) we rewrite the Eq. (4) in terms
of the projection operators:
\begin{equation}
H = \left( {\sum\limits_k {\left| k \right\rangle \left\langle k
\right|} } \right)H\left( {\sum\limits_m {\left| m \right\rangle
\left\langle m \right|} } \right) = \sum\limits_k {\varepsilon _k
\left| k \right\rangle \left\langle k \right|}  + {\bf{E}}\left( t
\right)\sum\limits_{k,m} {{\bf{d}}_{km} \left| k \right\rangle
\left\langle m \right|} ,
\end{equation}
where ${\bf{d}}_{km}  = \left\langle k \right|-e{\bf{r}}\left| m
\right\rangle $ is the matrix element of optical dipole transition
between the states $\left| {k} \right\rangle $ and $\left| {m}
\right\rangle $. The state vector of the system may be presented
in the form
\begin{equation}
\left| {\Psi \left( t \right)} \right\rangle  = \sum\limits_n {c_n
\left( t \right)e^{ - i\varepsilon _n t} \left| n \right\rangle }
\end{equation}
and is governed by the non-stationary Schr\"odinger equation
\begin{equation}
i\frac{{\partial \left| {\Psi \left( t \right)} \right\rangle
}}{{\partial t}} = H\left| {\Psi \left( t \right)}
 \right\rangle,
\end{equation}
with the initial condition $\left| {\Psi \left( t_0 \right)}
\right\rangle  = \alpha _0 \left| 0 \right\rangle  + \beta _0
\left| 1 \right\rangle $ (hereafter $\hbar  \equiv 1$).

Inserting Eqs. (5) and (6) into Eq. (7) we arrive at the set of
linear differential equations for the probability amplitudes $c_n
\left( t \right)$. We shall only examine the transitions between
the states $\left| {0} \right\rangle $ and $\left| {1}
\right\rangle $ and the states $\left\{ {\left| {k} \right\rangle
} \right\}_{exc}$:
\begin{equation}
\left\{ \begin{array}{l}
 i\dot {c_0}  = {\bf{E}}\left( t \right)\sum\limits_k {{\bf{d}}_{0k} c_k e^{ - i\omega _{0k} t} }  \\
 i\dot {c_1}  = {\bf{E}}\left( t \right)\sum\limits_k {{\bf{d}}_{1k} c_k e^{ - i\omega _{1k} t} }  \\
 i\dot {c_k}  = {\bf{E}}\left( t \right)\left( {{\bf{d}}_{0k}^* c_0 e^{i\omega _{0k} t}  + {\bf{d}}_{1k}^* c_1 e^{i\omega _{1k} t} } \right),\,\,k \in \left\{ {k} \right\}_{exc}, \\
 \end{array} \right.
\end{equation}
where $\omega _{0\left( 1 \right)k}  = \varepsilon _k  -
\varepsilon _{0\left( 1 \right)} $.

Let the electromagnetic field imposed on the structure to have (in
the dipole approximation) the form of two phase-locked pulses
\begin{equation}
{\bf{E}}\left( t \right) = {\bf{E}}_0 \left( t \right)\cos \left(
{\omega _0 t + \varphi _0 } \right) + {\bf{E}}_1 \left( t
\right)\cos \left( {\omega _1 t + \varphi _1 } \right),
\end{equation}
where the pulse envelopes ${\bf{E}}_0 \left( t \right) =
{\bf{E}}_0 f_0 \left( t \right),\,\,{\bf{E}}_1 \left( t \right) =
{\bf{E}}_1 f_1 \left( t \right)$ are the slowly-varying (compared
to optical frequencies) time-dependent functions, $\omega _{0,1} $
are the pulse frequencies, and $\varphi _{0,1} $ are the pulse
phases. We require both pulses to be in the two-photon resonance
with the DD structure, i.e. $\varepsilon_0 +  \omega_0 =
\varepsilon_1 +  \omega_1$ or, alternatively,
\begin{equation}
\delta _{0k}  = \delta _{1k}  \equiv \delta _k,
\end{equation}
where $\delta _{0\left( 1 \right)k}  = \omega _{0\left( 1 \right)}
- \omega _{0\left( 1 \right)k} $ is the detuning of the pulse
frequency $\omega _{0\left( 1 \right)} $ from the resonant
frequency $\omega _{0\left( 1 \right)k} $.

Making use of the rotating-wave approximation we obtain from Eqs.
(8) the following set:
\begin{equation}
\left\{ \begin{array}{l}
 i\dot c_0  = \sum\limits_k {\left[ {\lambda _{0k} \left( t \right) + \mu _{1k} \left( t \right)e^{i\Delta t} } \right]c_k e^{i\delta _k t} }  \\
 i\dot c_1  = \sum\limits_k {\left[ {\mu _{0k} \left( t \right)e^{ - i\Delta t}  + \lambda _{1k} \left( t \right)} \right]c_k e^{i\delta _k t} }  \\
 i\dot c_k  = \left[ {\lambda _{0k}^* \left( t \right) + \mu _{1k}^* \left( t \right)e^{ - i\Delta t} } \right]c_0 e^{ - i\delta _k t}  + \\ \,\,\,\,\,\,\,\,\ + \left[ {\mu _{0k}^* \left( t \right)e^{i\Delta t}  + \lambda _{1k}^* \left( t \right)} \right]c_1 e^{ - i\delta _k t} ,\,\,k \in \left\{ {k} \right\}_{exc}, \\
 \end{array} \right.
\end{equation}
where $\lambda _{0\left( 1 \right)k} \left( t \right) =\lambda
_{0\left( 1 \right)k}f_{0\left( 1 \right)}\left( t
\right)e^{i\varphi _{0\left( 1 \right)} } ,\,\,\mu _{0\left( 1
\right)k} \left( t \right) = \mu _{0\left( 1 \right)k}f_{0\left( 1
\right)}\left( t \right)e^{i\varphi _{0\left( 1 \right)} }$,
$\lambda _{0\left( 1 \right)k}= {{{\bf{d}}_{0\left( 1 \right)k}
{\bf{E}}_{0\left( 1 \right)} } \mathord{\left/
 {\vphantom {{{\bf{d}}_{0\left( 1 \right)k} {\bf{E}}_{0\left( 1 \right)} } 2}} \right.
 \kern-\nulldelimiterspace} 2}$, $\mu _{0\left( 1 \right)k} = {{{\bf{d}}_{1\left( 0 \right)k} {\bf{E}}_{0\left( 1 \right)}  } \mathord{\left/
 {\vphantom {{{\bf{d}}_{1\left( 0 \right)k} {\bf{E}}_{0\left( 1 \right)} } 2}} \right.
 \kern-\nulldelimiterspace} 2}$  and
the identities $\omega _{0\left( 1 \right)}  - \omega _{1\left( 0
\right)k}  = \delta _{1\left( 0 \right)k}  \mp \Delta $ are used.

Eqs. (11) describe the dynamical process involving many
three-level excitation schemes that act in parallel. Each of them
is characterized by the set of parameters $\Delta$, $\lambda
_{0\left( 1 \right)k},\,\mu _{0\left( 1 \right)k} $, and
$\delta_k$, where $k \in \left\{ {k} \right\}_{exc}$. We shall
suppose the values of $\lambda_{0(1)k}$ and $\mu_{0(1)k}$ to be of
the same order. Depending on the ratios between these parameters
$k$-th excitation scheme may be classified in the following way.
First we consider the case of small detunings. If the coupling
coefficients of the optical dipole transitions $\lambda_{0(1)k}$
and the detunings $\delta_k$ satisfy the inequality
\begin{equation}
\left| {\delta _k}\right| ,\Delta  \ll \left| {\lambda _{0k} }
\right|,\left| {\lambda _{1k} } \right|,
\end{equation}
the three-level scheme works in the {\it resonant symmetric}
regime. Instead, the applicability of the {\it resonant
asymmetric} scheme \cite{16} is provided by the condition
\begin{equation}
\left| {\delta _k}\right| \ll \left| {\lambda _{0k} }
\right|,\left| {\lambda _{1k} } \right| \ll \Delta.
\end{equation}
We see that the asymmetry/symmetry of the structure isn't defined
by the presence/absence of the energy difference $\Delta$ only but
by the ratio between $\Delta$ and the coupling coefficients
$\left| {\lambda _{0k} } \right|,\left| {\lambda _{1k} } \right|$
as well. In other words, a driven DD structure can be treated
(relative to the $k$-th transition scheme) as symmetric if the
influence of the parameter $\Delta$ introducing a "static"
asymmetry is compensated by an appropriate value of the field
strength defined from (12). In this case only one external pulse
is sufficient to excite both transitions \cite{18}.

Next we shall examine the opposite case where the states ${\left|
0 \right\rangle ,\left| 1 \right\rangle }$ are connected through
the off-resonant transitions involving the manyfold of the excited
states lying at the edge of the potential barrier. Two situations
are possible again, i.e.
\begin{equation}
\Delta  \ll \left| {\lambda _{0k} } \right|,\left| {\lambda _{1k}
} \right| \ll \left| {\delta _k}\right|
\end{equation}
 and
\begin{equation}
\left| {\lambda _{0k} } \right|,\left| {\lambda _{1k} } \right|
\ll \left| {\delta _k}\right| ,\Delta.
\end{equation}
The first of these inequalities corresponds to the {\it
off-resonant symmetric} excitation scheme. This situation was
studied in Refs. \cite{19,20} for the double-dot structures. As it
was shown, the set of single-qubit operations produced by such
one-electron dynamics is incomplete since in order to realize an
arbitrary rotation of the qubit-state vector the structure
symmetry must be broken. Here our attention will be focused on the
{\it off-resonant asymmetric} case for which the conditions (15)
are satisfied and each pulse drives the transitions between only
one of localized state $\left| {0} \right\rangle $ ($\left| {1}
\right\rangle $) and the transport states $\left\{ {\left| {k}
\right\rangle } \right\}_{exc}$. This implies also that the values
of $\Delta$ and $\delta_k$ must be rather different from each
other for all $k$ to prevent the single-donor resonant dynamics.
It may be attained, e.g., by setting $\omega_0 < \omega_{exc}$.

\centerline{\bf III. THE OFF-RESONANT DYNAMICS}

 Using the inequalities (15) let us average each of equations (11)
over the time interval $T_\Delta = {{2\pi } \mathord{\left/
 {\vphantom {{2\pi } \Delta }} \right.
 \kern-\nulldelimiterspace} \Delta }$ on which all of the time-dependent functions except $\exp \left( { \pm i\Delta t} \right)$
may be replaced by their mean values \cite{21} so that after
integration we arrive at the following set of equations:
\begin{equation}
\left\{ \begin{array}{l}
 i\dot {c_0}  = \sum\limits_k {\lambda _{0k} \left( t \right) b_k }  \\
 i\dot {c_1}  = \sum\limits_k {\lambda _{1k} \left( t \right) b_k }  \\
 i\dot {b_k} = - \delta _k b_k  + \lambda _{0k}^* \left( t \right)c_0  + \lambda _{1k}^* \left( t \right)c_1 ,\,\,k \in \left\{ {k} \right\}_{exc}, \\
 \end{array} \right.
\end{equation}
where $b_k  = c_k \exp \left( {  i\delta _k t} \right),\,k \in
\left\{ {k} \right\}_{exc} $. The applicability of Eqs. (16)
requires the pulse switching times $\tau_{0(1)sw}$ to be rather
long as compared with $T_\Delta$. The inequalities (15) allow one
to apply the adiabatic elimination procedure \cite{22} to the
intermediate levels $\left\{ {\left| {k} \right\rangle }
\right\}_{exc}$:
\begin{equation}
\dot {b_k } \approx 0,\,\,b_k  \approx {{\left[ {\lambda _{0k}^*
\left( t \right)c_0  + \lambda _{1k}^* \left( t \right)c_1 }
\right]} \mathord{\left/
 {\vphantom {{\left[ {\lambda _{0k}^* \left( t \right)c_0  + \lambda _{1k}^* \left( t \right)c_1 } \right]} {\delta _k }}} \right.
 \kern-\nulldelimiterspace} {\delta _k }},\,\,k \in \left\{ {k} \right\}_{exc}
\end{equation}
and equations for two remaining probability amplitudes $c_0,c_1$
 in the matrix form read
\begin{equation}
i\frac{\partial }{{\partial t}}\left( {\begin{array}{*{20}c}
   {c_0 }  \\
   {c_1 }  \\
\end{array}} \right) = \left( {\begin{array}{*{20}c}
   {\Lambda _{0} \left( t \right)} & {\Lambda _{2} \left( t \right)}  \\
   {\Lambda _{2}^* \left( t \right)} & {\Lambda _{1} \left( t \right)}  \\
\end{array}} \right)\left( {\begin{array}{*{20}c}
   {c_0 }  \\
   {c_1 }  \\
\end{array}} \right),
\end{equation}
where $\Lambda _{0} \left( t \right) = \Lambda _{0} f_0^2 \left( t
\right)$, $\Lambda _{1} \left( t \right) = \Lambda _{1} f_1^2
\left( t \right)$, $\Lambda _{2} \left( t \right) = \Lambda _{2}
 f_0 \left( t
\right)f_1 \left( t \right)$ and $\Lambda _{0}  = \sum\limits_k
{{{\left| {\lambda _{0k} } \right|^2 } \mathord{\left/
 {\vphantom {{\left| {\lambda _{0k} } \right|^2 } {\delta _k }}} \right.
 \kern-\nulldelimiterspace} {\delta _k }}} $,
$\Lambda _{1}  = \sum\limits_k {{{\left| {\lambda _{1k} }
\right|^2 } \mathord{\left/
 {\vphantom {{\left| {\lambda _{1k} } \right|^2 } {\delta _k }}} \right.
 \kern-\nulldelimiterspace} {\delta _k }}} $,
$\Lambda _{2}  = e^{i\left( {\varphi _0  - \varphi _1 }
\right)}\sum\limits_k {{{\lambda _{0k} \lambda _{1k}^* }
\mathord{\left/
 {\vphantom {{\lambda _{0k} \lambda _{1k}^* } {\delta _k }}} \right.
 \kern-\nulldelimiterspace} {\delta _k }}} $.

The eigenstates and the eigenenergies of the matrix in right-hand
side of Eq. (18) may be written as
\begin{equation}
\left\{ \begin{array}{l}
 \left|  +  \right\rangle  = e^{i\arg \left[ {\Lambda _{2} \left( t \right)} \right]} \cos \left[ {{{\Theta \left( t \right)} \mathord{\left/
 {\vphantom {{\Theta \left( t \right)} 2}} \right.
 \kern-\nulldelimiterspace} 2}} \right]\left| 0 \right\rangle  + \sin \left[ {{{\Theta \left( t \right)} \mathord{\left/
 {\vphantom {{\Theta \left( t \right)} 2}} \right.
 \kern-\nulldelimiterspace} 2}} \right]\left| 1 \right\rangle  \\
 \left|  -  \right\rangle  = e^{i\arg \left[ {\Lambda _{2} \left( t \right)} \right]} \sin \left[ {{{\Theta \left( t \right)} \mathord{\left/
 {\vphantom {{\Theta \left( t \right)} 2}} \right.
 \kern-\nulldelimiterspace} 2}} \right]\left| 0 \right\rangle  - \cos \left[ {{{\Theta \left( t \right)} \mathord{\left/
 {\vphantom {{\Theta \left( t \right)} 2}} \right.
 \kern-\nulldelimiterspace} 2}} \right]\left| 1 \right\rangle  \\
 \end{array} \right.
\end{equation}
and
\begin{equation}
E_ \pm  \left( t \right) = {{\left[ {\Lambda _{0} \left( t \right)
+ \Lambda _{1} \left( t \right)} \right]} \mathord{\left/
 {\vphantom {{\left[ {\Lambda _{0} \left( t \right) + \Lambda _{1} \left( t \right)} \right]} 2}} \right.
 \kern-\nulldelimiterspace} 2} \pm \Omega \left( t \right),
\end{equation}
respectively, where
\begin{equation}
\cos \left[ {\Theta \left( t \right)} \right] = {{\left[ {\Lambda
_{0} \left( t \right) - \Lambda _{1} \left( t \right)} \right]}
\mathord{\left/
 {\vphantom {{\left[ {\Lambda _{0} \left( t \right) - \Lambda _{1} \left( t \right)} \right]} {2\Omega \left( t \right)}}} \right.
 \kern-\nulldelimiterspace} {2\Omega \left( t \right)}},\,\,\,\,\sin \left[ {\Theta \left( t \right)} \right] = {{\left| {\Lambda _{2} \left( t \right)} \right|} \mathord{\left/
 {\vphantom {{\left| {\Lambda _{2} \left( t \right)} \right|} {\Omega \left( t \right)}}} \right.
 \kern-\nulldelimiterspace} {\Omega \left( t \right)}},
\end{equation}
and
\begin{equation}
\Omega \left( t \right) = \sqrt {{{\left[ {\Lambda _{0} \left( t
\right) - \Lambda _{1} \left( t \right)} \right]^2 }
\mathord{\left/
 {\vphantom {{\left[ {\Lambda _{0} \left( t \right) - \Lambda _{1} \left( t \right)} \right]^2 } 4}} \right.
 \kern-\nulldelimiterspace} 4} + \left| {\Lambda _{2} \left( t \right)} \right|^2 }
\end{equation}
is the instantaneous Rabi frequency. Using the unitary
transformation
\begin{equation}
D\left( t \right) = \left( {\begin{array}{*{20}c}
   {e^{i\arg \left[ {\Lambda _{2} \left( t \right)} \right]} \cos \left[ {{{\Theta \left( t \right)} \mathord{\left/
 {\vphantom {{\Theta \left( t \right)} 2}} \right.
 \kern-\nulldelimiterspace} 2}} \right]} & {e^{i\arg \left[ {\Lambda _{2} \left( t \right)} \right]} \sin \left[ {{{\Theta \left( t \right)} \mathord{\left/
 {\vphantom {{\Theta \left( t \right)} 2}} \right.
 \kern-\nulldelimiterspace} 2}} \right]}  \\
   {\sin \left[ {{{\Theta \left( t \right)} \mathord{\left/
 {\vphantom {{\Theta \left( t \right)} 2}} \right.
 \kern-\nulldelimiterspace} 2}} \right]} & { - \cos \left[ {{{\Theta \left( t \right)} \mathord{\left/
 {\vphantom {{\Theta \left( t \right)} 2}} \right.
 \kern-\nulldelimiterspace} 2}} \right]}  \\
\end{array}} \right)
\end{equation}
we represent the state vector in the instantaneous basis $\left\{
{\left|  +  \right\rangle ,\left|  -  \right\rangle } \right\}$ as
\begin{equation}
\left| \Phi \left( t \right)  \right\rangle  = a_ + \left( t
\right) \left|  + \right\rangle  + a_ - \left( t \right) \left|  -
\right\rangle ,\,\,\,\left| \Psi \left( t \right) \right\rangle =
D\left( t \right)\left| \Phi \left( t \right) \right\rangle
\end{equation}
and rewrite Eq. (18) in the new basis as
\begin{equation}
i\frac{\partial }{{\partial t}}\left( {\begin{array}{*{20}c}
   {a_ +  }  \\
   {a_ -  }  \\
\end{array}} \right) = \left( {\begin{array}{*{20}c}
   {E_ +  } & {{{\dot \Theta } \mathord{\left/
 {\vphantom {{\dot \Theta } 2}} \right.
 \kern-\nulldelimiterspace} 2}}  \\
   { - {{\dot \Theta } \mathord{\left/
 {\vphantom {{\dot \Theta } 2}} \right.
 \kern-\nulldelimiterspace} 2}} & {E_ -  }  \\
\end{array}} \right)\left( {\begin{array}{*{20}c}
   {a_ +  }  \\
   {a_ -  }  \\
\end{array}} \right),
\end{equation}
where
\begin{equation}
\dot \Theta \left( t \right) = \frac{{\left| {\left[ {\dot \Lambda
_0 \left( t \right) - \dot \Lambda _1 \left( t \right)}
\right]\left| {\Lambda _2 \left( t \right)} \right| - \left| {\dot
\Lambda _{2} \left( t \right)} \right|\left[ {\Lambda _0 \left( t
\right) - \Lambda _1 \left( t \right)} \right]} \right|}}{{\left[
{\Lambda _0 \left( t \right) - \Lambda _1 \left( t \right)}
\right]^2 /4  + \left| {\Lambda _2 \left( t \right)} \right|^2 }}.
\end{equation}
Here we restrict our interest by the diagonal evolution followed
from the choice of the system parameters for which $\dot \Theta
\left( t \right) \ll E_{\pm} \left( t \right)$. In this case the
solution of Eq. (25) is straightforward:
\begin{equation}
\begin{array}{l}
 a_ +  \left( t \right) = a_ +  \left( {t_0 } \right)\exp \left[ { - i\int\limits_{t_0 }^t {E_ +  \left( {t'} \right)dt'} } \right], \\
 a_ -  \left( t \right) = a_ -  \left( {t_0 } \right)\exp \left[ { - i\int\limits_{t_0 }^t {E_ -  \left( {t'} \right)dt'} } \right]. \\
 \end{array}
\end{equation}
With the help of equations (23), (24), and (27) we may write down
the expression for the evolution matrix of the qubit-state vector
$\left| {\Psi \left( t \right)} \right\rangle $ in the laboratory
frame:

\begin{equation}
\begin{array}{l}
 \left| {\Psi \left( t \right)} \right\rangle  = U\left| {\Psi \left( t_0 \right)} \right\rangle , \\
 U = \left( {\begin{array}{*{20}c}
   {e^{ - i\varepsilon _0 t} } & 0  \\
   0 & {e^{ - i\varepsilon _1 t} }  \\
\end{array}} \right)D\left( t \right)\left( {\begin{array}{*{20}c}
   {e^{ - i\int\limits_{t_0 }^t {E_ +  \left( {t'} \right)dt'} } } & 0  \\
   0 & {e^{ - i\int\limits_{t_0 }^t {E_ -  \left( {t'} \right)dt'} } }  \\
\end{array}} \right)D^\dag  \left( t_0 \right) =  \\
 \,\,\,\,\,\, = e^{ - i\left[ {\varepsilon _0 t + \varphi _\Lambda  \left( t \right)} \right]} \left( {\begin{array}{*{20}c}
   {u_{00} } & {u_{01} }  \\
   {u_{10} e^{ - i\Delta t} } & {u_{11} e^{ - i\Delta t} }  \\
\end{array}} \right), \\
 \end{array}
\end{equation}
where
\begin{equation}
\begin{array}{l}
 u_{00}  = u_{11}^*  = e^{ - i\tilde \Omega \left( t \right)} \cos \left[ {{{\Theta \left( t \right)} \mathord{\left/
 {\vphantom {{\Theta \left( t \right)} 2}} \right.
 \kern-\nulldelimiterspace} 2}} \right]\cos \left[ {{{\Theta \left( {t_0 } \right)} \mathord{\left/
 {\vphantom {{\Theta \left( {t_0 } \right)} 2}} \right.
 \kern-\nulldelimiterspace} 2}} \right] + e^{i\tilde \Omega \left( t \right)} \sin \left[ {{{\Theta \left( t \right)} \mathord{\left/
 {\vphantom {{\Theta \left( t \right)} 2}} \right.
 \kern-\nulldelimiterspace} 2}} \right]\sin \left[ {{{\Theta \left( {t_0 } \right)} \mathord{\left/
 {\vphantom {{\Theta \left( {t_0 } \right)} 2}} \right.
 \kern-\nulldelimiterspace} 2}} \right], \\
 u_{01}  =  - u_{10}^*  = \\
 =e^{i\arg \left[ {\Lambda _{2} \left( t \right)} \right]} \left\{ {e^{ - i\tilde \Omega \left( t \right)} \cos \left[ {{{\Theta \left( t \right)} \mathord{\left/
 {\vphantom {{\Theta \left( t \right)} 2}} \right.
 \kern-\nulldelimiterspace} 2}} \right]\sin \left[ {{{\Theta \left( {t_0 } \right)} \mathord{\left/
 {\vphantom {{\Theta \left( {t_0 } \right)} 2}} \right.
 \kern-\nulldelimiterspace} 2}} \right] - e^{i\tilde \Omega \left( t \right)} \sin \left[ {{{\Theta \left( t \right)} \mathord{\left/
 {\vphantom {{\Theta \left( t \right)} 2}} \right.
 \kern-\nulldelimiterspace} 2}} \right]\cos \left[ {{{\Theta \left( {t_0 } \right)} \mathord{\left/
 {\vphantom {{\Theta \left( {t_0 } \right)} 2}} \right.
 \kern-\nulldelimiterspace} 2}} \right]} \right\}, \\
 \end{array}
\end{equation}
and
\begin{equation}
\tilde \Omega \left( t \right) = \int\limits_{t_0 }^t {\Omega
\left( {t'} \right)dt'} ,\,\,\,\varphi _\Lambda  \left( t \right)
= \int\limits_{t_0 }^t {{{\left[ {\Lambda _{0} \left( {t'} \right)
+ \Lambda _{1} \left( {t'} \right)} \right]} \mathord{\left/
 {\vphantom {{\left[ {\Lambda _{0} \left( {t'} \right) + \Lambda _{1} \left( {t'} \right)} \right]} 2}} \right.
 \kern-\nulldelimiterspace} 2}dt'} .
\end{equation}
The expressions (28) - (30) describe the effective two-level
dynamics that corresponds to the continuous evolution of the qubit
state vector on the Bloch sphere. In the next section we show how
to choose the pulse and structure parameters in order to realize
the most important single-qubit gates.

\centerline{\bf IV. SINGLE-QUBIT OPERATIONS}

We illustrate the qubit state engineering by considering a
particular case of the driving pulses sharing the same time
dependence, - i.e. $f_0 \left( t \right) = f_1 \left( t \right)
\equiv f\left( t \right)$. The condition $\dot \Theta  = 0$ is
then satisfied and the components of the evolution matrix (28)
take the form
\begin{equation}
\begin{array}{l}
 u_{00}  = u_{11}^*  = \cos \left[ {\tilde \Omega \left( t \right)} \right] - i\cos \left( {\Theta _0 } \right)\sin \left[ {\tilde \Omega \left( t \right)} \right], \\
 u_{01}  =  - u_{10}^*  =  - ie^{i\arg \left( {\Lambda _{2} } \right)} \sin \left( {\Theta _0 } \right)\sin \left[ {\tilde \Omega \left( t \right)} \right], \\
 \end{array}
\end{equation}
where $\Theta _0  = \arcsin \left[ {{{\left| {\Lambda _{2} }
\right| } \mathord{\left/
 {\vphantom {{\left| {\Lambda _{2} } \right| } {\sqrt {{{\left( {\Lambda _{0}  - \Lambda _{1} } \right)^2 } \mathord{\left/
 {\vphantom {{\left( {\Lambda _{00}  - \Lambda _{11} } \right)^2 } 4}} \right.
 \kern-\nulldelimiterspace} 4} + \left| {\Lambda _{2} } \right|^2 } }}} \right.
 \kern-\nulldelimiterspace} {\sqrt {{{\left( {\Lambda _{0}  - \Lambda _{1} } \right)^2 } \mathord{\left/
 {\vphantom {{\left( {\Lambda _{0}  - \Lambda _{1} } \right)^2 } 4}} \right.
 \kern-\nulldelimiterspace} 4} + \left| {\Lambda _{2} } \right|^2 } }}} \right]$.
The dynamics described by the equations (31) is sufficient to
generate an arbitrary single-qubit rotation on the Bloch sphere.
For example, the quantum operations such as NOT ($\sigma _ x$):
$\left(\alpha_0, \beta_0 \right)^T\rightarrow \left(\beta_0,
\alpha_0 \right)^T$; PHASE ($\sigma _ z$): $\left(\alpha_0,
\beta_0 \right)^T\rightarrow \left(\alpha_0, -\beta_0 \right)^T$;
and Hadamard ($H$): $\left(\alpha_0, \beta_0 \right)^T\rightarrow
\left[ \left( \alpha_0 + \beta_0 \right)/ \sqrt{2}, \left(
\alpha_0 - \beta_0 \right)/ \sqrt{2}  \right]^T$ can be realized
(up to the common phase) given the following choices of the pulse
- structure parameters:
\begin{equation}
\tilde \Omega \left( T \right) = {\pi  \mathord{\left/
 {\vphantom {\pi  {2 + \pi k,\,\,\,}}} \right.
 \kern-\nulldelimiterspace} {2 + \pi k,\,\,\,}}T\Delta  = 2\pi l,\,\,\,\arg \left( {\Lambda _{2} } \right) = 2\pi m
\end{equation}
and
\begin{equation}
\Theta_0 \left( {\sigma _x } \right) = {\pi  \mathord{\left/
 {\vphantom {\pi  2}} \right.
 \kern-\nulldelimiterspace} 2} + \pi n,\,\,\,\Theta_0 \left( {\sigma _z } \right) = \pi n,\,\,\,\Theta_0 \left( H \right) = {\pi  \mathord{\left/
 {\vphantom {\pi  4}} \right.
 \kern-\nulldelimiterspace} 4} + \pi n,
\end{equation}
respectively. Here $k,l,m,$ and $n$ are the integers and $T$ is
the pulse duration. Of course, this is not a unique parameter
choice to attain the above quantum operations. Let us evaluate the
time $T_{NOT}$ needed to implement NOT operation. For the pulse
strengths $E_0 \sim E_1 \sim 10$ V/cm and $a_B \sim 3$ nm one has
$|\lambda_{0k}|\sim|\lambda_{1k}|\sim ea_B E_0\sim10^{-5}$ eV. By
setting $\delta_k \sim - 10^{-4}$ eV, $k=k_{exc}^{min}$ that
ensures the validity of Eqs. (14) and (15) for all $k$, we obtain
$|\Lambda_0|\geq |\lambda_{0k}|^2 /|\delta_k|\sim 10^{9}$
s$^{-1}$. Hence, $T_{NOT}\sim\Lambda_0^{-1}$ is of order of
nanoseconds. More careful estimation requires the detailed
knowledge of the energy spectrum of DD structure and the values of
$\lambda_{0(1)k}$.

Note that the complete population transfer between the qubit
states, or NOT
 operation, requires that $\Lambda _{0}  = \Lambda _{1}$. This
is naturally met for nearly symmetric DD structures where $\Delta
\approx 0$ and $\left| {{\bf{d}}_{0k} } \right| \approx \,\left|
{{\bf{d}}_{1k} } \right|$. In general, however, one should keep in
mind that $\left| {{\bf{d}}_{0k} } \right| \ne \,\left|
{{\bf{d}}_{1k} } \right|$ that makes the performing of the
condition $\Lambda _{0}  = \Lambda _{1}$ very problematic. It
seems then reasonable to point the other way for the population
transfer based upon the pulse-shaped techniques. Such methods,
e.g., the stimulated Raman adiabatic passage (STIRAP) \cite{23},
are very robust against the pulse/structure imperfections and
would allow one to handle the quantum information carefully. The
theory of the adiabatic population transfer via multiple
intermediate states, including the off-resonant case, was
presented in Ref. \cite{24}. Note that for the pulses strongly
detuned from the resonance, the time ordering is no more important
since successful population transfer may be attained for both
intuitive and counterintuitive pulse sequences. If initially $c_0
\left( t_0 \right) = 1,\,c_1 \left( t_0 \right) = 0$, the
intuitive (counterintuitive) pulse ordering means that $\mathop
{\lim }\limits_{t \to  t_0 } \left[ {{{f_0 \left( t \right)}
\mathord{\left/
 {\vphantom {{f_0 \left( t \right)} {f_1 \left( t \right)}}} \right.
 \kern-\nulldelimiterspace} {f_1 \left( t \right)}}} \right] = \infty \left( 0
 \right)$ and $\mathop {\lim }\limits_{t \to T } \left[ {{{f_0 \left( t
\right)} \mathord{\left/
 {\vphantom {{f_0 \left( t \right)} {f_1 \left( t \right)}}} \right.
 \kern-\nulldelimiterspace} {f_1 \left( t \right)}}} \right] = 0\left( \infty
 \right)$ and, as it follows from Eq. (21), $\Theta \left( t_0 \right) = 0\left( \pi
 \right)$, $\Theta \left( T  \right) = \pi \left( 0 \right)$.
 The population transfer may be understood as the adiabatic temporal development of the
eigenstate $\left| +  \right\rangle$ ($\left| -  \right\rangle$)
for the intuitive (counterintuitive) pulse ordering. As it is seen
from Eq. (29), the qubit state inversion is realized in the
asymmetric DD structures if the conditions $\arg \left[ {\Lambda
_2 \left( T \right)} \right] \pm \tilde \Omega \left( T \right) =
\pi n$ and $T\Delta = \pi \left( {2m + 1} \right)$ are fulfilled.
The detailed analysis concerning the arrangement of the pulse
shapes in STIRAP can be found elsewhere \cite{23}.

\centerline{\bf V. DISCUSSION }

To provide more clarity in the understanding of the advantages of
the charge qubit-state engineering presented above, let us compare
the resonant and off-resonant excitation schemes. The resonant
optical driving of the DD structure modelled by single three-level
scheme has been studied in Refs. \cite{16,17}. In the case
considered here a more complex dynamics can take place involving
more than one three-level scheme. For example, if we tune the
lasers on resonance with the transition between the logical
subspace $\left\{ {\left| 0 \right\rangle ,\left| 1 \right\rangle
} \right\}$ and a state $\left| {r} \right\rangle ,\,\,r \in
\left\{ {k } \right\}_{exc}$ located near the top of the barrier,
a number of states with the energies close to $\varepsilon _{r} $
will be excited as well. This picture is quite expected in the
hydrogen-like molecular ions for the high-lying states which
energies are within the interval $\Delta \varepsilon _{r} \le
\left| {\lambda _{0(1)r} } \right|$. However, for the symmetric
structure it means that there will be no transitions between the
qubit states. It is because the exited states belonging to the
same doublet are presented by the symmetric and antisymmetric
superpositions of the excited states of isolated donors which,
 being excited simultaneously, interfere constructively on one donor
  and destructively on another one. As a consequence, the donors
are excited independently. This effect becomes more significant as
the interdonor distance $R$ increases and the tunnel coupling
between the donors decreases. When the energy splitting of the
maximally-resolved doublet becomes comparable with the coupling
coefficients of the optical dipole transitions, the process of the
electron transfer between the donors is terminated. Note that the
optically driven DD structure will demonstrate the similar
behavior if one of the pulses is short enough so that its duration
is $T \le {1 \mathord{\left/
 {\vphantom {1 {\Delta \varepsilon _{r} }}} \right.
 \kern-\nulldelimiterspace} {\Delta \varepsilon _{r} }}$ and thus
 it contains harmonics in the frequency range $\delta \omega  \sim {1 \mathord{\left/
 {\vphantom {1 T}} \right.
 \kern-\nulldelimiterspace} T} \ge \Delta \varepsilon _{r} $.
 Again, the states with the energies belonging to the interval $\Delta \varepsilon _{r}$ will be
 excited simultaneously giving rise to the electron transfer
 blockage just outlined.

The reliable resonant scheme thus deals with single transport
state and is very sensitive to the pulse detuning from the
resonance with that state. For example, the non-zero detuning
always produces an amplitude error in NOT gate because of
incomplete depopulation of the initial state when the pulse is off
\cite{18}. On the contrary, the use of the off-resonant pulses
enables one to exploit the whole number of excited states.
Moreover, we don't need to control the pulse frequencies  with
high accuracy since a small variation in the pulse detunings
brings about an insignificant change in the Rabi frequency (30).
The computational errors originated from the frequency
renormalization can be corrected by the corresponding change in
the pulse duration due to the smooth time dependencies of the
probability amplitudes. The only requirement that must be followed
closely for successful electron state manipulations is the Raman
two-photon resonant condition (10).

The selectivity of the electron resonant transfer requires also a
strict control over the pulse polarizations. The transport states
in the molecular ion are formed through the hybridization of those
individual donor states whose orbitales are extended along the
axis $x$ that coincides with the interdonor direction. Other
states are hybridized weakly and cannot assist efficiently in the
electron dynamics. Their excitations are due to the pulse
components polarized along the axes $y$ and $z$. It amounts to the
population leakage into the non-hybridized single-donor states
with the energies lying in the close proximity to the energy of
the transport state. Let us define the small angles
$\gamma_{n\,y}$ and $\gamma_{n\,z}$ that characterize the
deviations of the $n$-th pulse polarization from the axis $x$:
\begin{equation}
\begin{array}{l}
 {\bf{E}}_n  = {\bf{E}}_{n\,x} + {\bf{E}}_{n\,y} \cos \left( {{\pi  \mathord{\left/
 {\vphantom {\pi  2}} \right.
 \kern-\nulldelimiterspace} 2} + \gamma _{n\,y} } \right) + {\bf{E}}_{n\,z} \cos \left( {{\pi  \mathord{\left/
 {\vphantom {\pi  2}} \right.
 \kern-\nulldelimiterspace} 2} + \gamma _{n\,z} } \right), \\
 \,\,\,\,\,\,\,\,\,\,\,\,\,\,\,\,\,\,\,\,\,\,\,\,\,\,\,\,\,\,\,\,\,\,\,\,\,\,\,\,\,\,\,\,\left| {\gamma _{n\,y} } \right|,\,\,\left| {\gamma _{n\,z} } \right| \ll 1,\,\,\,\,\,n = 0,1 ,\\
 \end{array}
\end{equation}
 then the probability of successful implementation of the quantum
 operations is reduced by a factor of $w \sim 1 - \max({\gamma _ {ny}^2,\gamma _ {nz}^2})$. In the off-resonant case, the
populations of those states remain negligibly small ($ \sim
{{\left| {\lambda _{n k} } \right|^2 } \mathord{\left/
 {\vphantom {{\left| {\lambda _{0,1\,k} } \right|^2 } {\delta _k^2 }}} \right.
 \kern-\nulldelimiterspace} {\delta _k^2 }}$) and the
corresponding channel of population leakage is blocked.

The important difference between the resonant and off-resonant
schemes lies in the treatment of the decoherence problem. We know
the relaxation rates from the transport state caused by the
spontaneous photon/phonon emission during the resonant excitation
\cite{17} may be high enough to corrupt the qubit state. In the
off-resonant scheme the population of the intermediate state(s) is
negligible and the probability of relaxation is drastically
reduced. The influence of the residual population of the
intermediate state on the adiabatic electron transfer in the
three-level scheme was examined in Ref. \cite{25} for the gaussian
pulses. It was shown that the error introduced by the spontaneous
emission together with the error due to the non-adiabaticity are
inversely proportional to the pulse detuning and can be made small
enough to allow the fault tolerant quantum computation.

In our model of the single-electron off-resonant transitions we
don't consider in detail the spectrum of the DD structure. The
 classification of excited states used as the transport
states as well as their wave functions are still to be determined.
Besides, the participation of the continuum states in the electron
dynamics has been completely ignored. These states, in principle,
can also be used as transport channels and, at the same time, can
bring about additional decoherence (see, e.g., \cite{26}). Both of
these issues will be the topics of further development of the
model.

\centerline{\bf VI. CONCLUSIONS }

In this paper we have considered the one-electron double-donor
structure subjected to the action of two off-resonant
electromagnetic pulses. Unlike the other systems proposed to serve
as the potential candidates for the solid state
optically-controlled qubits (double quantum dots, rf-SQUIDs), the
double-donor structure is characterized by sufficiently high
density of the bound states at the edge of the barrier that
separates the donors. It means that the three-level resonant
scheme proposed earlier to implement the desired qubit-state
evolution may be unsuitable to maintain the appropriate
selectivity of the optical excitations. On the other hand, the
off-resonant scheme seems to be more efficient for the qubit
manipulations and robust in comparison with the resonant scheme.
Though the Raman evolution of the qubit is slower than that in the
case of the resonant driving, it seems to be more reliable for the
realization of quantum operations. We have shown that the basic
single-qubit operations may be performed on the DD structure for
several pulse and structure parameter choices. The use of the
adiabatic schemes is of the particular interest.

It is also very important that we can operate on the qubit without
the detailed knowledge of the spectral properties of the DD
structure. Keeping the frequencies of the detuned pulses smaller
than the difference between the energy of the lowest state of the
excited manifold and the energy of the qubit state $\left| 1
\right\rangle $, we automatically use all states from that
manifold as the transport states. The information about the
structure and pulse parameters is contained in the Rabi frequency
of the two-level oscillations. This frequency can be defined
experimentally for each set of the detunings, the strengths, and
the durations of the pulses. The results of those measurements
could be used to reconstruct the features of the spectrum of the
DD structure.

Note that the method of the electron-state manipulations by
optical means can be applied also to the spin-based encoding
schemes like that of Ref. \cite{2}. The implementation of
optically controlled effective electron spin exchange described in
Ref. \cite{27} for the two-electron double-dot structure, can be
generalized on the two-electron DD structure.

\centerline{\bf ACKNOWLEDGMENTS }

Discussions with L. A. Openov are gratefully acknowledged.

\newpage

Fig. 1. Schematics of the quantum state manipulation in the
one-electron double-donor structure. The qubit states $\left| 0
\right\rangle $ and $\left| 1 \right\rangle $ are defined by the
localized orbital states of the donors $A$ and $B$ with the
energies $\varepsilon _ 0$ and $\varepsilon _ 1$, respectively.
They are coupled by two optical pulses with the frequencies
$\omega_0$ and $\omega_1$ being in the two-photon resonance
($\varepsilon_0 +  \omega_0 = \varepsilon_1 +  \omega_1$). The
pulses are detuned from the excited (delocalized over the
structure) state $\left| k \right\rangle$ by the detuning
$\delta_k$. The energy difference $\Delta=\varepsilon _ 1 -
\varepsilon _ 0$ is introduced by the
 voltage $V>0$ applied to the gate on the left of the
 structure.

\newpage

\includegraphics[width=\hsize]{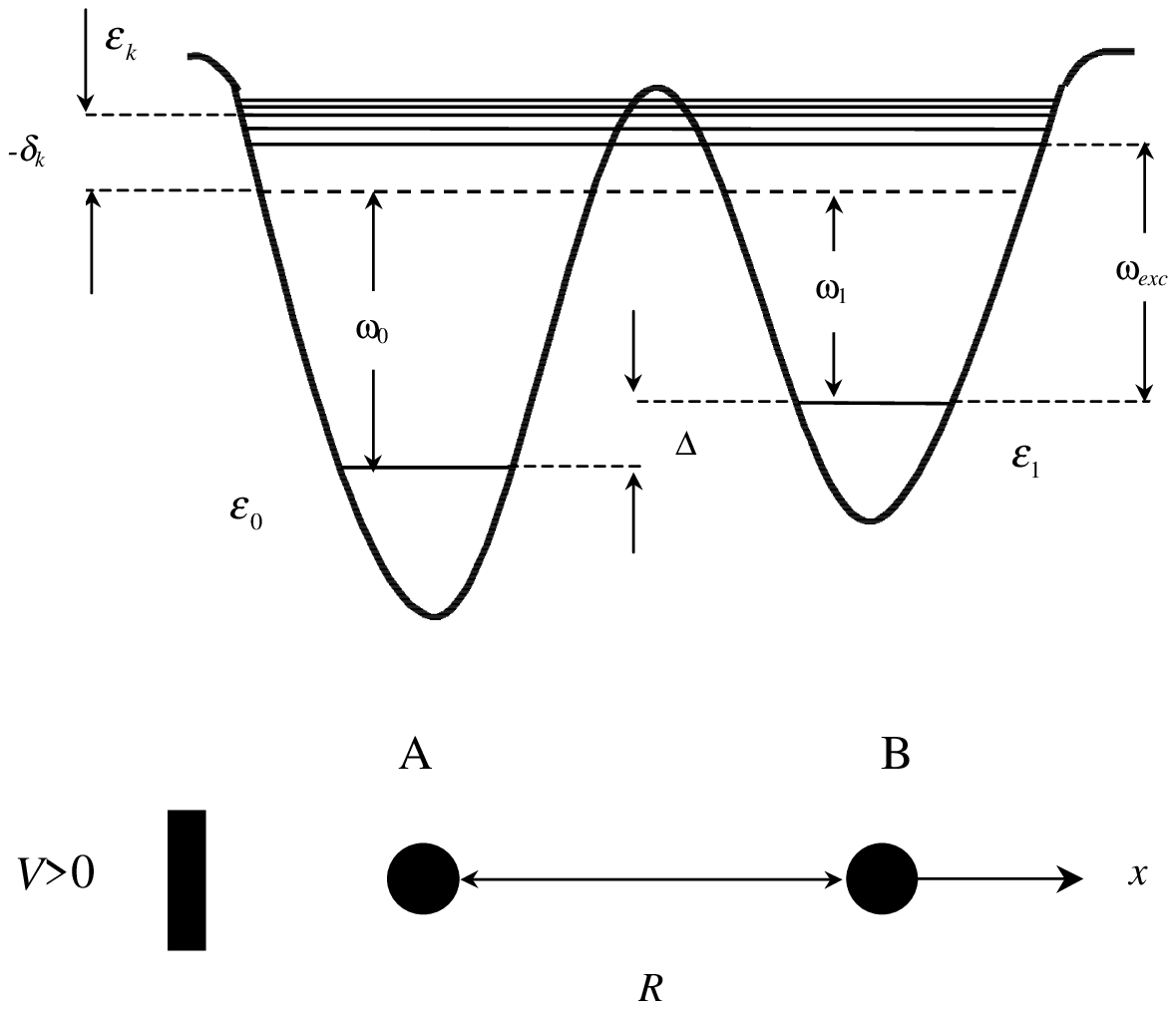}

\vskip 6mm

\end{document}